\documentclass[preprint,preprintnumbers,amsmath,amssymb]{revtex4}
\usepackage{graphicx}
\usepackage{dcolumn}
\usepackage{bm}
\begin{document}
\draft
\title{Quantum gravitational optics in the field of a gravitomagnetic monopole}
\author{ N.~Ahmadi$^{a}$\footnote{
Electronic address:~nahmadi@ut.ac.ir}, S.~Khoeini-Moghaddam$^{b}$\footnote{Electronic
address:~saloumeh@mehr.sharif.edu} and M.~Nouri-Zonoz $^{a,c}$ \footnote{
Electronic address:~nouri@theory.ipm.ac.ir}}
\address{$^{a}$ Department of Physics, University of Tehran, North Karegar Ave., Tehran 14395-547, Iran. \\
$^{b}$Department of Physics, Sharif
University of Technology, P O Box 11365-9161 Tehran, Iran. \\
$^{c}$ Institute for studies in theoretical physics and mathematics, P O Box 19395-5531 Tehran, Iran.}
\begin{abstract}
Vacuum polarization
in QED in a background gravitational field induces interactions which 
{\it effectively}
modify the classical picture of light rays as the null geodesics of spacetime. After a short introduction on the main aspects of the quantum gravitational optics, as a nontrivial example, we study this effect in the background of NUT space characterizing the spacetime of a spherical mass endowed with a gravitomagnetic monopole charge, the so called NUT factor.
\end{abstract}
\pacs{PACS No., .Nr}
\maketitle
\section{Introduction}
It is not unexpected that the effect of QED vacuum polaraization in a background gravitational field, though tiny, should modify the photon propagation in that field. Vacuum polarization is an
effect in which the photon exists as a virtual $e^{+}e^{-}$ pair
for a short time. This virtual transition assigns photons with an effective size of
${\mathcal{O}}(\lambda_c)$, where $\lambda_c$ is
the Compton wavelength of the electron \cite{Drummond}-\cite{shore}. 
Obviously the photon
propagation will be affected by the gravitational field if the scale of the spacetime curvature $L$ is comparable to $\lambda_c$.
The influence of the vacuum polarization on the propagation of a bundle of rays has been studied recently through the perturbative deformation of the Raychaudhuri equation \cite{ahm}. Ignoring the vacuum polarization, the equivalence principle leads to the photon propagation at the speed of
light on the spacetime null geodesics. But when the pair production is taken into account,
the equivalence principle is violated in such a way that the
photon trajectories are modified and superluminal (photon)
propagation, without necessarily breaking down the causality \cite{shore}, becomes a possibility.
\subsection{Quantum gravitational optics}
Considering the vacuum polarization, the first order correction to the electromagnetic action
\begin{equation}
W_0=-\frac{1}{4}\int d^4x\sqrt{-g}F_{\mu\nu}F^{\mu\nu}
\end{equation}
is given by
\begin{eqnarray}\label{Drumhath}
W_1&=&\frac{1}{m_{e}^{2}}\int
d^4x(-g)^{\frac{1}{2}}(aRF_{\mu\nu}F^{\mu\nu}+bR_{\mu\nu}F^{\mu\sigma}F^{\nu}_{\sigma}+cR_{\mu\nu\sigma\tau}F^{\mu\nu}F^{\sigma\tau}
+dD_{\mu}F^{\mu\sigma}D_{\nu}F^{\nu}_{\sigma})
\end{eqnarray}
in which 
\begin{eqnarray}\label{coefficients}
a=-\frac{1}{144}\frac{\alpha}{\pi}\hspace{1.5cm}b=\frac{13}{360}\frac{\alpha}{\pi}\hspace{1.5cm}c=-\frac{1}{360}\frac{\alpha}{\pi}
\hspace{1.5cm}d=-\frac{1}{30}\frac{\alpha}{\pi}
\end{eqnarray}
are perturbative coefficients of ${\mathcal{O}}(\alpha)$, $m_{e}$ is the electron mass and $\alpha$ is the fine structure constant. Expression (\ref{Drumhath}) is called the {\em Drummond-Hathrell} action \cite{Drummond}.
The first three terms represent the influence of curvature while the last one
exists even in the flat space-time an shows the off-mass-shell
effects in the vacuum polarization \footnote {The coefficients $a$, $b$ and $c$
may be obtained from the coupling of a graviton to two
on-mass-shell photons in the flat-space limit. The term with
coefficient {\it d} is neglected in the following study because it is of
${\mathcal{O}}(\alpha^2)$.}. The effective equation of motion can be obtained by varying $W=W_0+W_1$ with respect to $A_\nu$ as follows,
\begin{equation}\label{equation of motion }
D_{\mu}F^{\mu\nu}-\frac{1}{m_{e}^{2}}[2bR_{\mu\sigma}D^{\mu}F^{\sigma\nu}+4cR_{\mu\hspace{3mm}\sigma\tau}^{\hspace{2mm}\nu}D^{\mu}F^{\sigma\tau}]=0
\end{equation}
in which only terms of first order in $\alpha$ and
$\left(\lambda_c/L\right)^{2}$ are kept.
Now we use the geometric optics approximation 
in which the electromagnetic field is written as a slowly-varying
amplitude and a rapidly-varying phase through a small
parameter $\epsilon$, i.e, $ (A_\mu+i\epsilon
B_\mu+\cdots)\exp(i\frac{\vartheta}{\epsilon})$, where $\vartheta$
is a real scalar field. In geometric optics, the wave vector is defined as the gradient of the phase, i.e, $k_\mu=
\partial_\mu\vartheta$ and the polarization vector is introduced through $ A_\mu=Aa_\mu$, where $A$ is the amplitude and $a_\mu$ is the polarization vector.
The geometric optics method gives the light cone condition at ${\mathcal{O}}(\epsilon^{-1})$ as \cite{shore}
\begin{eqnarray}\label{light cone}
k^2-\frac{2b}{m_{e}^{2}}R_{\mu\nu}k^\mu
k^\nu+\frac{8c}{m_{e}^{2}}R_{\mu\nu\sigma\tau}k^\mu k^\sigma a^\nu
a^\tau=0,
\end{eqnarray}
in which $a^\mu a_\mu = -1$. Since this effective equation of motion is homogeneous and quadratic in $k^\mu$ it can be written as
\begin{equation}
{\mathcal G^{\mu\nu}}k_\mu k_\nu=0.
\end{equation}
 In other words at this order of approximation there is no dispersion
 and consequently the phase and group velocities are equal and given by
\begin{equation}
v_{ph}=\frac{k^0}{|k|}
\end{equation}
Now there is a possibility that the modified null cone lies outside the original light cone leading to superluminal photon velocities.
The photon trajectories can be deduced from the modified equation of motion as
\cite{shore95}
\begin{eqnarray}\label{geodesic}
\frac{d^2x^{\nu}}{ds^2}+\Gamma^{\nu}_{\sigma\tau}\frac{dx^\sigma}{d
s}\frac{dx^\tau}{d s}-
\frac{1}{m_{e}^{2}}D^\nu\left[(bR_{\sigma\tau}-4cR_{\sigma\rho\tau\lambda}a^\rho
a^\lambda)\frac{dx^\sigma}{d s}\frac{dx^\tau}{d s}\right]=0.
\end{eqnarray}
In the classical general relativity the photons move along
spacetime geodesics which are the integral curve of $k^\mu=\frac{dx^\mu}{d s}$. This is no longer the case when quantum corrections are taken
into account due to the modification produced by the third term in eq. (\ref{geodesic}). For Schwarzschild space-time, it is
shown that the paths depend on the
polarization of the photon and are given by classical trajectories plus a correction term which depends on the black hole mass,
the electron mass and the photon energy.
\section{Quantum corrections in NUT space}
The NUT spacetime line element is given by \cite{NUT}:
\begin{equation}\label{NUT}
ds^2=e^{-2\nu}[dt-2q(1+\cos\theta)d\phi]^2-(1-\frac{q^2}{r^2})^{-1}e^{2\nu}dr^2-r^2d\theta^2-r^2\sin^2\theta
d\phi^2
\end{equation}
where
\begin{equation}
e^{-2\nu}=1-\frac{2}{r^2}(q^2+M\sqrt{r^2-q^2})
\end{equation}
This is a stationary spherically symmetric
solution of the vacuum Einstein field
equations with two parameters, the mass $M$ and the NUT factor $q$. Obviously when $q$ vanishes, the metric reduces to that of the Schwarzschild spacetime. It is more convenient to introduce a local orthonormal frame with the following basis
1-forms 
\begin{eqnarray}
e^0&=&e^0_{\hspace{2mm}t}dt+e^0_{\hspace{2mm}\phi}d\phi\nonumber\\
e^1&=&e^1_{\hspace{2mm}r}dr\nonumber\\
e^2&=&e^2_{\hspace{2mm}\theta}d\theta\\
e^3&=&e^3_{\hspace{2mm}\phi}d\phi\nonumber
\end{eqnarray}
where the vierbeins are
\begin{eqnarray}\label{vierbeins}
e^0_{\hspace{2mm}t}&=&e^{-\nu}\hspace{1cm}\nonumber\\
e^0_{\hspace{2mm}\phi}&=&-2qe^{-\nu}(1+\cos\theta)\nonumber\\
e^1_{\hspace{2mm}r}&=&1/\sqrt{1-q^2/r^2}e^{\nu}\\
e^2_{\hspace{2mm}\theta}&=&r\nonumber\\
e^3_{\hspace{2mm}\phi}&=&r\sin\theta\nonumber
\end{eqnarray}
Calculating the connection one forms it could be seen that there are nine non-vanishing components of Riemann tensor out of which the following
three are independent  
\begin{equation}\label{defination1}
R_{0101}\equiv A\hspace{1cm}R_{0202}\equiv
B\hspace{1cm}R_{0312}\equiv C
\end{equation}
where
\begin{eqnarray}\label{definitions}
A&=&\left(1-\frac{q^2}{r^2}\right)
e^{-2\nu}\left(\nu''-2\nu'^2+\frac{q^2}{r^3(1-\frac{q^2}{r^2})}\nu'\right)\nonumber\\
B&=&r^{-1}e^{-2\nu}\left(\nu'(1-\frac{q^2}{r^2})-q^2r^{-3}\right)\nonumber\\
C&=&qr^{-3}\sqrt{1-\frac{q^2}{r^2}}e^{-2\nu}\left(1+r\nu'\right)
\end{eqnarray}
Introducing the following notation \cite{Drummond}-\cite{shore}
\begin{equation}
U^{ij}_{ab}=\delta^{i}_{a}\delta^{j}_{b}-\delta^{i}_{b}\delta^{j}_{a}\;\;\;\;\; i,j=0,1,2,3,\nonumber
\end{equation}
the Riemann tensor can be written compactly as:
\begin{eqnarray}\label{Riemann Tensor}
R_{abcd}=&2&A(U^{01}U^{01}-U^{23}U^{23})+2B(U^{02}U^{02}+U^{03}U^{03}-U^{12}U^{12}-U^{13}U^{13})\\\nonumber
&+&C(U^{03}U^{12}+U^{12}U^{03}-U^{02}U^{13}-U^{13}U^{02}
-2U^{01}U^{23}-2U^{23}U^{01}).
\end{eqnarray}
To keep the expression clear we have suppressed the lower indices {\it a, b, c, d} in all terms.
For a Ricci flat spacetime, the light cone
condition, eq. (\ref{light cone}), is given by:
\begin{eqnarray}\label{photon propagation}
k^2a_b+\epsilon R_{abcd}k^a k^c a^d=0.
\end{eqnarray}
where $\epsilon=-\frac{8c}{m_{e}^{2}}$. This is a set of three
simultaneous linear equations for independent components of the
polarization $a_b$. It is convenient to introduce the three
independent linear combinations of momentum components as \cite{Drummond};
\begin{equation}\label{definition02}
l_b=k^aU^{01}_{ab}\hspace{2cm}m_b=k^aU^{02}_{ab}\hspace{2cm}n_b=k^aU^{03}_{ab}
\end{equation}
together with three dependent combinations
\begin{eqnarray}\label{defination2}
p_b&=&k^aU^{12}_{ab}=\frac{1}{k^0}(k^2m_b-k^3l_b)\\\nonumber
q_b&=&k^aU^{13}_{ab}=\frac{1}{k^0}(k^2n_b-k^3l_b)\\\nonumber
r_b&=&k^aU^{23}_{ab}=\frac{1}{k^0}(k^2n_b-k^3m_b)
\end{eqnarray}
The vectors $l$, $m$ and $n$ are independent and orthogonal to
$k^a$.
To solve photon propagation equation, we rewrite eq.
(\ref{photon propagation}) as a set of equations for the independent
polarization components $a.l$, $a.m$ and $a.n$ by substituting for the Riemann tensor from eq. (\ref{Riemann Tensor}). Doing so we arrive at;
\begin{eqnarray}\label{set of equations}
k^2a.l&+&2\epsilon A\left(l^2l.a\right)+2\epsilon
B\left(m.lm.a+n.ln.a-p.lp.a-q.lq.a-r.lr.a \right)\nonumber\\
&+&\epsilon
C\left(n.lp.a+p.ln.a-m.lq.a-q.lm.a-2l^2r.a-2r.ll.a\right)=0\nonumber\\
k^2a.m&+&2\epsilon A\left( l.ml.a\right)+2\epsilon B\left(
m^2m.a+n.mn.a-p.mp.a-q.mq.a-r.mr.a\right) \nonumber\\
&+&\epsilon
C\left(n.mp.a+p.mn.a-m^2q.a-q.mm.a-2l.mr.a-2r.ml.a\right)=0\nonumber\\
k^2a.n&+&2\epsilon A\left(l.nl.a\right)+2\epsilon
B\left(m.nm.a+n^2n.a-p.np.a-q.nq.a-r.nr.a \right)\nonumber\\
&+&\epsilon C\left(
n^2p.a+p.nn.a-m.nq.a-q.nm.a-2l.nr.a-2r.nl.a\right)=0
\end{eqnarray}
Although difficult, in principle it is possible to solve the above system of coupled equations in general.
We select special trajectories which illustrate the most important
features.
\subsection{Radial component of motion }
For the radial component of motion in which the
photon momentum components satisfy $k^2=k^3=0$ eq. (\ref{set of equations}) could be written in the following matrix form;
\begin{equation}\label{matrix}
\left(%
\begin{array}{ccc}
 k^2+2\epsilon A(-k^0k^0+k^1k^1) & 0 & 0 \\
  0 & k^2+2\epsilon B(-k^0k^0+k^1k^1)& 0 \\
  0 & 0 & k^2+2\epsilon B(-k^0k^0+k^1k^1) \\
  \end{array}%
  \right)
  \left(%
\begin{array}{c}
  l.a \\
  m.a \\
  n.a \\
\end{array}%
\right) =0
\end{equation}
 One can solve for the eigenvectors of the above matrix as the polarization 
vectors satisfying (\ref{photon propagation}). The $k^{2}$ value for each polarization is obtained through the vanishing of the determinant of (\ref{matrix}). Since $\epsilon $ is a small parameter \footnote{$\epsilon$ is of order $\frac{\alpha}{\lambda_c}$ but $A\epsilon$, $B\epsilon$ and $C\epsilon$ are of small order ${\alpha}{\left(\lambda_c/L\right)^2}$.}
we keep only the terms up to ${\cal O}(\epsilon)$ and therefore all the second terms in the diagonal components are ignored so that the only degenerate  eigenvalue is $k^2=0$. 
This is a  result expected from the spherical symmetry of NUT space and it means
there are no corrections to the radial component of motion and in consequence, the  photon velocity is given by $|\frac{k^0}{k^1}|=1$ for all the polarizations.
\subsection{Orbital component of motion}
Now consider the orbital component of motion in which the photon polarization is such that $k^1=k^2=0$. 
Diagonalizing the corresponding matrix we find as eigenvalues $\lambda_0 = k^2$ corresponding to the unphysical polarization vector, $a_{b}$, proportional to $n_{b}$ and
\begin{equation}
\lambda_{\pm} = k^2 \pm \epsilon\left[9C^{2}(k^0k^3)^2+(A-B)^2\left(k^0k^0+k^3k^3\right)^2\right]^{\frac{1}{2}}
\end{equation}
corresponding to the transverse polarization eigenvectors. From this the velocity shift is found to be
\begin{eqnarray}\label{velocity shift}
\delta v=\pm \frac{3}{2}\frac{M}{r^3}\epsilon\left[1+\frac{1}{4}\frac{1}{M^2}\left(\frac{q}{r}\right)^{2}\left(r^2-27M^2\right)+\frac{9}{2}\frac{1}{M}\left(\frac{q}{r}\right)^{2}\sqrt{r^2-q^2}+\frac{9}{4}\frac{1}{M^2}\left(\frac{q}{r}\right)^{4}\left(3r^2+8M^2\right)\right.\nonumber\\
\left.-\frac{24}{M}\left(\frac{q}{r}\right)^{4}\sqrt{r^2-q^2}-\frac{6}{M^2}\left(\frac{q}{r}\right)^{6}\left(3r^2+M^2\right)-\frac{24}{M}\left(\frac{q}{r}\right)^{6}\sqrt{r^2-q^2}\right]^{\frac{1}{2}}
.\end{eqnarray}
The pure mass term, $\frac{3}{2}\frac{M}{r^3}\epsilon=\frac{1}{30}\frac{\alpha}{\pi}\frac{1}{m_{e}^{2}}\frac{M}{r^3}$ is factored out to see the contribution of the NUT factor in the velocity shift as compared to that of mass.
\section{The trajectories}
To obtain the new trajectories the modification term in equation (\ref{geodesic}) should be calculated along with the classical approximation to $O\left(\alpha^{0}\right)$. Although these equations are too complicated to be useful in general, employing the space-time symmetries will help us to obtain useful information about the quantum modifications. In this respect we note that
the third term in (\ref{geodesic}) is a derivative of a scalar and so the covariant derivative can be replaced by a partial derivative. On the other hand, as mentioned before, this scalar is polarization dependent but since it is calculated along the classical paths it possesses all the classical symmetries. In space-times with spherical symmetry such as  NUT, Schwartzschild and Reissner-Nordstrom, it depends only on the radial coordinate $r$ and so the only modified component is the $r$-component. Employing the (1+3) threading formulation of space-time decomposition \cite{Lyn} to NUT space it could be written in the following general form;
 \begin{equation}
 ds^{2}=e^{-2\nu}\left(dt-A_{\alpha}dx^{\alpha}\right)^{2}-\gamma_{\alpha\beta}
dx^{\alpha}dx^{\beta}.\label{2}\end{equation}
 In the classical case varying $x^{\alpha}$, we find
 \begin{eqnarray}
 \delta x^{\alpha}\left\{\frac{d}{d\tau}\left[e^{-2\nu}
\left(\dot{t}-A_{\beta}\dot{x}^{\beta}\right)A_{\alpha}+\gamma_{\alpha\beta}\dot{x}^{\beta}\right]+\frac{1}{2}\frac{\partial}{\partial x^{\alpha}}\left[e^{-2\nu}\left(\dot{t}-A_{\beta}\dot{x}^{\beta}\right)^{2}-\gamma_{\beta\gamma}\dot{x}^{\beta}\dot{x}^{\gamma}\right]\right\}=0
 .\label{3}\end{eqnarray}
 Here the projected 3-dimensional metric $\gamma_{\alpha\beta}$ can be 
written in the form involving the unit Cartesian vector $\hat{r}$
  \begin{equation}
 \gamma_{\alpha\beta}dx^{\alpha}dx^{\beta}=e^{2\lambda}dr^2+r^{2}
\left(d\hat{r}\right)^2.\label{4}\end{equation} 
where $e^{2\lambda}\equiv (1-\frac{q^2}{r^2})^{-1}e^{2\nu} $ and $\delta \hat{r}$
the variation of $\hat{r}$ is an arbitrary small 
vector perpendicular to $\hat{r}$. Now making variations with respect to $\hat{r}$ and $r$ and inserting the modification term along $r$, we obtain \cite{Lyn},
  \begin{equation}
\frac{d}{d\tau}\left(r^2 \frac{d\hat{r}}{d\tau}\right)+2\kappa\frac{d\hat{r}}{d\tau}\times q\hat{r}=0
  ,\label{5}\end{equation}
and
  \begin{equation}
   \kappa^2 e^{2\nu}-\dot{r}^2 e^{2\lambda}-L^2 r^{-2}+\frac{1}{m_{e}^{2}}
\frac{\partial}{\partial r}\left(4cR_{\beta\sigma\gamma\tau}a^{\sigma}a^{\tau}\frac{dx^{\beta}}{d\tau}\frac{dx^{\gamma}}{d\tau}\right)=0
  ,\label{6}\end{equation}
  where $L=r\times \frac{d}{d\tau}$ and 
$\kappa= e^{-2\nu}\left(\dot{t}-A_{\alpha}dx^{\alpha}\right)$.
   Crossing both sides of eq. (\ref{5}) by $\hat{r}$,
  \begin{equation}
 \hat{r}\times\frac{d}{d\tau}
\left(r^2 \frac{d\hat{r}}{d\tau}\right)=\frac{dL}{dt}=-2\frac{d}{d\tau}
\left(\kappa q\hat{r}\right)
  .\label{7}\end{equation}
    we end up with
   \begin{equation}
 L+2\kappa q\hat{r}=J={const}
  .\label{8}\end{equation}
i.e. the cone underlying the classical null trajectories is not modified by the quantum corrections. In classical limit, to see the geometry of the trajectory, we introduce the curvilinear angle $\phi$ around the cone's surface. Then $r^{2}\dot{\phi}=L$, so eq. (\ref{NUT}) can be integrated by quadrature
  \begin{equation}
 \phi_{c}-\phi_{0}=\int{\frac{Lr^{-2}dr}
{\sqrt{\kappa^{2}e^{-2(\lambda-\nu)}-L^{2}r^{-2}e^{-2\lambda}}}}
  .\label{9}\end{equation}
 In general this integral can not be performed explicitly for $\lambda$ 
and $\nu$ of NUT space, however, it has a $\phi_{c}\left(r\right)$ solution, which solved for $r$ gives $ r_{c}\left(\phi\right)$. The corresponding quantum modified trajectory is given by
  \begin{eqnarray}
\phi-\phi_{0}&=&\int{\frac{Lr^{-2}dr}{\sqrt{\kappa^{2}e^{-2(\lambda-\nu)}-L^{2}r^{-2}e^{-2\lambda}+e^{-2\lambda}f\left(r,a\right)}}},\nonumber\\  
 f\left(r,a\right)&= &\frac{\partial}
{\partial r}\left(4cR_{\beta\sigma\gamma\tau}a^{\sigma}a^{\tau}\frac{dx^{\beta}}{d\tau}
\frac{dx^{\gamma}}{d\tau}\right)
  .\label{10}\end{eqnarray}
with a solution in the form of $\phi=\phi_{c}\left(r\right)+\phi_{1}\left(r,a\right)$ which can be solved to give $r=r_{c}\left(\phi\right)+r_{1}\left(\phi,a\right)$. Trajectory modifications induced by vacuum polarization are along the $r$-component for the two physical polarizations, equal in magnitude but opposite in sign \footnote{This can be seen in the proof of the polarization sum rule theorem \cite{shore95}.}. The two modified trajectories, corresponding to two transverse polarizations, on the $r-\phi$ plane, are expected to be shifted in opposite directions with respect to the classical path such that the difference from the classic path varies with the value of the $r$ coordinate at each point on the path. We recall that the trajectory splitting happens on the classical cone and reduce to the space-time null geodesics when $\alpha \rightarrow 0$. The similar situation happens in Schwartzchild and Reiner-Nordestrom cases. For Schwartzchild space-time, the corresponding shift for circular trajectories has been demonstrated in \cite{thesis}. 
\section*{Acknowledgment}
N. A and M. N-Z thank University of Tehran for supporting this project 
under the grants provided by the research council.

\smallskip

\medskip
\begin{thebibliography}{15}
\bibitem[1]{Drummond}I.T. Drummond and S.J. Hathrell 1980, {\it Phys. Rev.} {\bf D 22}, 343
\bibitem[2]{shore}G. M. Shore 2003 {\it Contemp. Phys.} {\bf 44}, 503--21 ({\it preprint} gr-qc/0304059)
\bibitem[3]{ahm} N. Ahmadi and M. Nouri-Zonoz 2006 {\it Phys. Rev.} {\bf D 74}, 044034 ({\it Preprint} gr-qc/0605009) 
\bibitem[4]{shore95}G. M. Shore 1996 {\it Nucl. Phys.} {\bf B 460}, 379 ({\it Preprint} qr-qc/9504041)
\bibitem[5]{NUT}E. Newman, L. Tamburino, T. Unti 1963 {\it J. Math. Phys.} {\bf 4}, 915--23
\bibitem[6]{Lyn}D. Lynden-Bell and M. Nouri-Zonoz 1998 {\it Rev. Mod. Phys.}{\bf 70}, 2, 427 ({\it Preprint} gr-qc/9612049)
\bibitem[7] {thesis}A. S. Sehra {\it preprint} 2006 gr-qc/0605147
\end{thebibliography}
\end{document}